On shape dependence of the toxicity of rutile nanoparticles

Martin Breza

Department of Physical Chemistry, Slovak Technical University, Radlinskeho 9, SK-81237 Bratislava, Slovakia

e-mail: martin.breza@stuba.sk

Abstract

Using B3LYP method, the nearly spheric structure of $[Ti_7O_{28}H_{26}]^{2-}$, the rod-like structures of $[Ti_2O_{10}H_{10}]^{2-}$ and $[Ti_7O_{30}H_{10}]^{2-}$ chains and the structures of their neutral complexes with $Cu^{2+}$ coordinated at various terminal or bridging oxygen sites were optimized in order to assess the toxicity of rutile nanoparticles. Cu – ligand interaction energy parameters and Cu charges were evaluated. Spheric structures are more reactive than the rod-like chains of the (nearly) same size. The reverse relation holds for the degree of their toxicity as indicated by the extent of the electron density transfer to a $Cu^{2+}$ probe. The experimentally observed higher cytotoxicity of rod-like nanoparticles in comparison with the spheric ones might be explained by the higher electron density transfer to the interacting cells.

Introduction

Biochemical and molecular mechanisms of cytotoxicity include oxidative stress-induced cellular events and alteration of the pathways pertaining to intracellular calcium homeostasis [1]. All the stresses lead to cell injuries and death. The physiochemical properties of nanoparticles influence how they interact with cells and, thus, their overall potential toxicity. Understanding these properties can lead to the development of safer nanoparticles.

Recent studies have begun identifying various properties that make some nanoparticles more toxic than others [1]. The properties of nanoparticles that contribute to cytotoxicity include, but are not limited to, surface, particle size, particle morphology, and dissolution of ions. As oxidative stress is elevated and intracellular calcium homeostasis is perturbed due to exposure to nanoparticles, subsequent actions lead to cell injury and death, and deregulation of the cell cycle.

i) Particle size is likely to contribute to cytotoxicity [1,2]. Given the same mass, smaller nanoparticles have a larger specific surface area and thus more available surface area to interact with cellular components such as nucleic acids, proteins, fatty acids, and carbohydrates. The smaller size also likely makes it possible to enter the cell, causing cellular damage.

ii) Particle surface charge may affect the cellular uptake of particles as well as how the particles interact with organelles and biomolecules. Consequently, particle surface charge influences cytotoxicity.

iii) Shape also affects levels of toxicity. Rod-shaped $Fe_2O_3$ nanoparticles were found to produce much higher cytotoxic responses than sphere-shaped $Fe_2O_3$ nanoparticles in a murine macrophage cell line (RAW 264.7) [3]. Rod-shaped $CeO_2$ nanoparticles were found to produce more toxic effects in RAW 264.7 cells than octahedron or cubic particles [4]. Nanorod ZnO particles are more toxic toward human lung epithelial cells (A549) than the corresponding spherical ones [5]. Why the physical shape of a nanoparticle influences cytotoxicity remains to be elucidated.

Theoretical model studies based on quantum-chemical calculations are capable to shed more light on these problems. The relation between activity of some antioxidants and their copper coordination ability based on B3LYP calculations their metal ion affinity (MIA) values has been investigated [6-8]. The stability order of the antioxidant ligands with metals bonded at various coordination sites strongly depends on their position and nature. The spin density of the $Cu^{2+}$ cation upon ligand coordination becomes vanishingly small, whereas the ligand spin density approaches 1. Thus, the ligand is oxidized to a radical cation (Ligand$^{\bullet+}$), while Cu(II) is reduced to Cu(I). In agreement with experimental investigations, the higher antioxidant activity of individual compounds and their reaction sites may be assigned to higher MIA values and higher reducing character toward Cu(II). Another modification of the above-mentioned method has been used for both N centers of a series of para-phenylene diamine (PPD) antioxidants [9]. Nearly linear dependence of the experimental antioxidant effectiveness on Cu(II)-PPD interaction energies, Cu atomic charges and other electron density parameters has been deduced. This method might be used to estimate the level of oxidation stress caused by nanoparticles and thus of their cytotoxicity.

Nano-sized $TiO_2$ particles can be found in a large number of foods, cosmetic goods and consumer products. Their nanotoxicity has been drawn an increasing attention because human bodies are potentially exposed to this nanomaterial either by inhalation, oral or dermal route. Numerous studies have tried to characterize their *in vivo* biodistribution, clearance and toxicological effects (see e.g. [10-16]).

Rutile is the most stable polymorph of $TiO_2$ at all temperatures exhibiting lower total free energy than the metastable phases of anatase or brookite [17]. Rutile has a tetragonal unit cell (space group $P4_2$/mnm) [18]. The titanium cations are surrounded by an octahedron of 6 oxygen atoms while the oxygen anions have a coordination number of 3. Rutile crystals are most commonly observed to exhibit a prismatic or acicular growth habit with preferential orientation along their c-axis, the [001] direction. This growth habit is favored as the {110} facets of rutile exhibit the lowest surface free energy and are therefore thermodynamically most stable [19].

The rutile (110)–aqueous solution interface structure was measured [20] in deionized water (DIW) at 25 °C by the X-ray crystal truncation rod method. The rutile surface consists of a stoichiometric (1 : 1) surface unit mesh with the surface terminated by bridging oxygen (BO) and terminal oxygen (TO) sites, with a mixture of water molecules and hydroxyl groups (OH) occupying the TO sites.

Very recently, scanning tunnelling microscopy and surface X-ray diffraction were used [21] to determine the structure of the rutile (110)–aqueous interface, which is comprised of an ordered array of hydroxyl molecules with molecular water in the second layer. A combination of experimental data with interpretation aided by DFT calculations implies that the rutile $TiO_2$ (110) surface has terminal hydroxyls in the contact layer. The ideal coverage by terminal OH groups is half a monolayer, but this is decreased to approximately 0.4 monolayers by absences at domain wall boundaries.

The liquids in human body are, in principle, aqueous solutions which implies a possible protonation of the negative charged surface of the rutile nanoparticles. For the sake of simplicity, only hexacoordinated Ti atoms and full protonation of non-bridging O atoms may be supposed in model systems. The aim of this study is to compare the toxicity of small idealized protonated rod-like and sphere-like rutile nanoparticles based on their Cu(II) complexation ability and electron density transfer to Cu at DFT level of theory.

Method

Geometries of the model systems under study in their lowest ground spin states (denoted by spin multiplicity as the left superscript) were optimized using B3LYP functional [22] with standard 6-311G* basis sets for all atoms [23 – 25]. Stability of the optimized structures was confirmed by vibrational analysis (no imaginary vibrations). Atomic charges were evaluated in terms of Mulliken population analysis (MPA) [26], atomic polar tensor (APT) derived charges [27] and Natural Population Analysis (NPA) [28]. All the calculations were performed using Gaussian09 program package [29].

The metal-ligand interaction energy $\Delta_{int}E$ is defined as

$$\Delta_{int}E = E_{Complex} - E_L - E_{ion} \qquad (1)$$

where $E_{Complex}$ and $E_L$ are the energies of the $[L...Cu]^{q+2}$ complex and of the isolated rutile nanoparticle $L^q$ model cluster in their optimized geometries, respectively, and $E_{ion}$ is the energy of the isolated $Cu^{2+}$ ion [6 - 9]. Metal-ligand interaction enthalpy $\Delta_{int}H_{298}$ and Gibbs free energy $\Delta_{int}G_{298}$ data at 298 K were evaluated analogously.

Results and discussion

In the first step it was necessary to build suitable model systems based on experimental rutile structure [18] with fully protonated surface oxygen atoms of nearly spheric and rod-like shapes which are small enough to be treated by DFT methods. This task is complicated by additional requirements on equal charges and (nearly) the same number of atoms in order to eliminate charge and size effects.

The nearly spherical structure of $^1[Ti_7O_{28}H_{26}]^{2-}$ (model A) in the singlet ground state consists of a single $TiO_6$ octahedron surrounded by six lateral $TiO_6$ octahedra (see Fig. 1 for the optimized geometry).

A rod-like chain structure consists of $TiO_6$ octahedra bonded by double μ-OH bridges as the single bridged structures disintegrate after protonation (into fragments consisting of mostly two $TiO_6$ units). For comparative purposes we investigate both the dimeric $^1[Ti_2O_{10}H_{10}]^{2-}$ (model B) and heptameric $^1[Ti_7O_{30}H_{10}]^{2-}$ (model C) chains in the singlet ground state (see Fig. 2 for the optimized geometries). It may be seen that only up to three $TiO_6$ units are bonded by double OH bridges after protonation and these trimers are bonded by the single ones only. The reason of such behavior is unknown.

In the next step we placed a $Cu^{2+}$ ion in the neighborhood of the oxygen atom (ca 1.7 Å distance) of various terminal (TO) hydroxyl (models A1, B1 and C1) or bridging (BO) hydroxyl (models A2, B2 and C2) groups. In this way optimized structures are presented in Figs. 3 - 6 and Tables 1 and 2. As our model systems contain many oxygen atoms, Cu atoms in the optimized structures are bonded at least to two O atoms and at least one of them is TO. In A model series Cu atoms are always bonded to BO whereas in B and C models the Cu-BO bonding is rare (see B2a and C2b models). In some cases $Cu^{2+}$ ions can cause splitting terminal hydroxyl groups (A1a and B1c models) or even $Ti(OH)_4$ units (C1b, C1d and C1e models). Adding $Cu^{2+}$ ions to TO or (more frequently) to BO sites can cause hydroxyl bridges splitting. No $Ti(OH)_4$ unit removal proceeds due to Cu – BO bond formation. As expected, in all optimized structures the Cu – BO bonds are longer than the Cu-TO ones whereas the shortest Cu-O bonds belong to free hydroxyls (A1a and B1c models).

Positive Cu charges in all model systems (Tables 1 and 2) are always significantly lower than +2 (i.e. the free $Cu^{2+}$ ion charge in all types of population analysis). The highest average Cu charge (i.e. the lowest electron density transfer from the ligand) is observed for nearly spherical A models whereas these ones for B and C models are nearly equal (i. e. within their standard deviations). The highest Cu charges are in model systems with the highest number of Cu-O bonds (A2a, B1c and C2b models). MPA charges are lower than the APT ones whereas the NBO ones are highest. Nevertheless, all these population analysis types preserve the same trends in Cu charges. In agreement with [6 - 8] only vanishing spin density remains at Cu atoms in optimized structures (not presented).

Metal-ligand interaction energies (Table 3) evaluated in terms of DFT energies and Gibbs free energies are practically equal whereas these ones in terms of enthalpies are lower. Nevertheless, in all three cases the same trends are preserved. Their average absolute values for model systems under study decrease in the sequence B > A > C. The highest energy is released for the systems with maximal numbers of Cu-O bonds (A2a and B1c models) or with a maximal number of broken OH bridges (C2c model). Coiled C chains around Cu atoms are important for the energy effects as well.

Conclusions

Our results show that the electron density removal from model rutile nanoparticles is connected with significant changes in their structure and may cause their degradation as well. Unfortunately, we cannot obtain structure with only one Cu – O bond but terminal hydroxyls seem to be more reactive than the bridging ones as indicated by the geometries and interaction energies of the optimized structures. Spheric structures are more reactive than the rod-like chains of the (nearly) same size. The reverse relation holds for the degree of their toxicity as indicated by the extent of the electron density transfer to a $Cu^{2+}$ probe. Experimental proof of these conclusion might be based on cytotoxicity comparison of rutile nanoparticles of various shapes (and of similar size distributions). Nevertheless, in the light of our results the higher toxicity of the above mentioned rod-shaped $Fe_2O_3$, $CeO_2$ or ZnO nanoparticles in comparison with the spheric ones [3 - 5] may explained by the higher electron density transfer to the interacting cells.

Although the scientific community has made considerable strides in understanding nanotoxicity in the recent past, the future research needed to decipher nanotoxicity remain significant [1]. For instance, what are the properties of the nanoparticle that induce oxidative stress? How do nanoparticles interact, physically and chemically, with biomolecules such as nucleic acids, proteins, and lipids leading to alteration of gene expression? The answer on the last question in [1] – what is the basic scientific principle that dictates the shape-dependent cytotoxicity? - might be based on the results of our study. We are convinced that quantum-chemical modeling is necessary for answering all these questions. As quantum-chemical calculations of larger model systems are connected with serious technical problem at DFT level of theory, an ONIOM treatment [30] combining DFT, semiempirical and molecular mechanics calculations must be used.

Acknowledgements

This project has received funding from the European Union's Horizon 2020 research and innovation programme under grant agreement No 685817 (HISENTS). We thank the HPC center at the Slovak University of Technology in Bratislava, which is a part of the Slovak Infrastructure of High Performance Computing (SIVVP Project ITMS 26230120002. funded by the European Region Development Funds) for computing facilities.

Appendix.

Supplementary data contain calculated energy terms of the systems under study and coordinates of optimized structures of $^1[Ti_7O_{28}H_{26}]^{2-}$ - model A, $^1[Ti_2O_{10}H_{10}]^{2-}$ - model B and $^1[Ti_7O_{30}H_{10}]^{2-}$ - model C.

References:

1. Yue-Wern Huang, Melissa Cambre and Han-Jung Lee, Int. J. Mol. Sci. 2017 , 18 , 2702 and the references therein

2. Jiang, J.; Oberdorster, G.; Elder, A.; Gelein, R.; Mercer, P.; Biswas, P. Does Nanoparticle Activity Depend upon Size and Crystal Phase? Nanotoxicology **2008**, 2, 33–42

3. Lee, J.H.; Ju, J.E.; Kim, B.I.; Pak, P.J.; Choi, E.K.; Lee, H.S.; Chung, N. Rod-shaped iron oxide nanoparticles are more toxic than sphere-shaped nanoparticles to murine macrophage cells. Environ. Toxicol. Chem. **2014**, 33, 2759–2766

4. Forest, V.; Leclerc, L.; Hochepie, J.F.; Trouvé, A.; Sarry, G.; Pourchez, J. Impact of Cerium Oxide Nanoparticles Shape on their In Vitro Cellular Toxicity. Toxicol. In Vitro **2017**, 38, 136–141.

5. I.-L. Hsiao, Y.-J. Huang / Science of the Total Environment 409 (2011) 1219–1228

6. Giuliano Alagona and Caterina Ghio *J. Phys. Chem. A* 2009, *113,* 15206–15216

7. Giuliano Alagona and Caterina Ghio Phys. Chem. Chem. Phys., 2009, 11, 776–790

8. Liliana Mammino J Mol Model (2013) 19:2127–2142

9. Puškárová I., Breza M.: DFT studies of the effectiveness of p-phenylenediamine antioxidants through their Cu(II) coordination ability.Polym. Degrad. Stab. 128, 15-21 (2016)

10. Olmedo D, Guglielmotti MB, Cabrini RL. An experimental study of the dissemination of Titanium and Zirconium in the body. Journal of Materials Science-Materials in Medicine. 2002; 13: 793–796.

11. Wang JX, Zhou GQ, Chen CY, Yu HW, Wang TC, Ma YM, et al. Acute toxicity and biodistribution of different sized titanium dioxide particles in mice after oral administration. Toxicology Letters. 2007; 168: 176–185

12. Fabian E, Landsiedel R, Ma-Hock L, Wiench K, Wohlleben W, van Ravenzwaay B. Tissue distribution and toxicity of intravenously administered titanium dioxide nanoparticles in rats. Archives of Toxicology. 2008; 82: 151–157.

13. Xie, G., Wang, C., Zhong, G., Toxicology Letters Volume 205, Issue 1, 2011, Pages 55-61

14. Wang Y, Chen Z, Ba T, Pu J, Chen T, Song Y, et al. Susceptibility of young and adult rats to the oral toxicity of titanium dioxide nanoparticles. Small. 2013; 9: 1742–1752.

15. Geraets L, Oomen AG, Krystek P, Jacobsen NR, Wallin H, Laurentie M, et al. Tissue distribution and elimination after oral and intravenous administration of different titanium dioxide nanoparticles in rats. Part Fibre Toxicol. 2014; 11: 30.

16. Elgrabli, D., Beaudouin, R., Jbilou, N., Floriani, M.[c], Pery, A., Rogerieux, F., Lacroix, G. PLoS ONE Volume 10, Issue 4, 2015, Article number e0124490

17. Hanaor, D. A. H.; Assadi, M. H. N.; Li, S.; Yu, A.; Sorrell, C. C. (2012). "Ab initio study of phase stability in doped TiO$_2$". Computational Mechanics. **50** (2): 185–94.

18. Diebold, Ulrike (2003). "The surface science of titanium dioxide" (PDF). Surface Science Reports. **48** (5–8): 53–229

19. Dorian A.H.Hanaor, Wanqiang Xu, Michael Ferry, Charles C.Sorrell Journal of Crystal Growth <u>Volume 359</u>, 15 November 2012, Pages 83-91

20. Zhan Zhang, Paul Fenter, Neil C. Sturchio, Michael J. Bedzyk, Michael L. Machesky, David J. Wesolowski Surface Science 601 (2007) 1129–1143

21. H. Hussain, G. Tocci1, T.Woolcot, X. Torrelles, C. L. Pang, D. S. Humphrey, C. M. Yim, D. C. Grinter1, G. Cabailh, O. Bikondoa, R. Lindsay, J. Zegenhagen, A. Michaelides and G. Thornton, *Nature Materials* volume 16, pages 461–466 (2017)

22. A. D. Becke, *J. Chem. Phys.*, **98** (1993) 5648-52.

23. D. McLean and G. S. Chandler, *J. Chem. Phys.*, **72** (1980) 5639-48.

24. K. Raghavachari, J. S. Binkley, R. Seeger, and J. A. Pople, *J. Chem. Phys.*, **72** (1980) 650-54.

25. A. J. H. Wachters, *J. Chem. Phys.*, **52** (1970) 1033.

26. Mulliken RS J. Phys. Chem. 23 (1955) 1833, 1841

27. J. Cioslowski, *J. Am. Chem. Soc.*, **111** (1989) 8333-36.

28. A. E. Reed, L. A. Curtiss, and F. Weinhold, *Chem. Rev.*, **88** (1988) 899-926.

29. Frisch MJ, Trucks GW, Schlegel HB, Scuseria GE, Robb MA, Cheeseman JR, et al., Gaussian 09, Revision D.01, Gaussian Inc., Wallingford, CT, 2009.

30. Dapprich S, Komáromi I, Byun KS, Morokuma K, Frisch MJ (1999) A New ONIOM Implementation in Gaussian 98. 1. The Calculation of Energies, Gradients and Vibrational Frequencies and Electric Field Derivatives. J Mol Struct (Theochem) 462: 1-21.

Table 1. MPA ($q(Cu)_{MPA}$), APT ($q(Cu)_{APT}$) and NBO ($q(Cu)_{NBO}$) copper atomic charges, the lengths of Cu-O bond to ($d_{Cu-OH}$) oxygen atoms of terminal (TO) and bridging (BO) hydroxyl groups in the model systems of A and B series.

| Compound | Model | $q(Cu)_{MPA}$ | $q(Cu)_{APT}$ | $q(Cu)_{NBO}$ | $d_{Cu-OH}$ [Å] |
|---|---|---|---|---|---|
| $^2[Ti_7O_{28}H_{26}Cu]^0$ | A1a | 0.518 | 0.829 | 0.957 | 1.764<br>1.958(TO)<br>1.996(BO)<br>1.902(TO)<br>1.967(TO) |
| | A2a | 0.684 | 1.054 | 1.048 | 1.970(TO)<br>2.026(BO) |
| | A2b | 0.403 | 0.541 | 0.710 | 1.935(TO)<br>1.958(BO) |
| | Average | 0.53±0.14 | 0.81±0.26 | 0.90±0.18 | |
| $^2[Ti_2O_{10}H_{10}Cu]^0$ | B1a | 0.406 | 0.585 | 0.719 | 1.976(TO) [a]<br>1.956(TO) [a] |
| | B1b | 0.286 | 0.507 | 0.629 | 1.918(TO)<br>1.923(TO) |
| | B1c | 0.540 | 0.959 | 1.001 | 1.818<br>1.939(TO) [a]<br>1.941(TO) [a] |
| | B2a | 0.330 | 0.565 | 0.710 | 2.074(TO) [a]<br>2.095(TO) [a]<br>2.154(BO) |
| | Average | 0.39±0.11 | 0.65±0.21 | 0.76±0.16 | |

[a] Terminal oxygen atoms bonded to the same Ti centre

Table 2. MPA (q(Cu)$_{MPA}$), APT (q(Cu)$_{APT}$) and NBO (q(Cu)$_{NBO}$) copper atomic charges, the lengths of Cu-O bond to (d$_{Cu-OH}$) oxygen atoms of terminal (TOi) and bridging (BOij) hydroxyl groups bonded to th i-th and j-th Ti centers (see Chart 1), respectively, and the indices of broken $\mu_{ij}$-OH bridges of the C model series.

| Compound | Model | q(Cu)$_{MPA}$ | q(Cu)$_{APT}$ | q(Cu)$_{NBO}$ | d$_{Cu-OH}$ [Å][a] | i - j |
|---|---|---|---|---|---|---|
| $^1$[Ti$_7$O$_{30}$H$_{30}$]$^{2-}$ | C | | | | | 1 − 2 |
| | | | | | | 4 − 5 |
| $^2$[Ti$_7$O$_{30}$H$_{30}$Cu]$^0$ | C1a | 0.382 | 0.632 | 0.711 | 1.951(TO1) | 1 − 2 |
| | | | | | 1.969(TO1) | 4 − 5 |
| | C1b | 0.408 | 0.579 | 0.734 | 2.003(TO2) | 1 − 2 (2x) |
| | | | | | 1.951(TO2) | 4 − 5 |
| | C1c | 0.368 | 0.546 | 0.659 | 1.986(TO1) | 1 − 2 |
| | | | | | 1.883(TO2) | 4 − 5 |
| | C1d | 0.379 | 0.533 | 0.630 | 1.888(TO2) | 1 − 2 (2x) |
| | | | | | 1.878(TO3) | 2 − 3 |
| | | | | | | 4 − 5 |
| | C1e | 0.314 | 0.520 | 0.652 | 1.886(TO3) | 1 − 2 (2x) |
| | | | | | 1.900(TO4) | 2 − 3 (2x) |
| | | | | | | 4 − 5 |
| | C1f | 0.437 | 0.533 | 0.686 | 1.885(TO6) | 1 − 2 |
| | | | | | 1.898(TO5) | 4 − 5 |
| | | | | | | 5 − 6 |
| | C1g | 0.343 | 0.513 | 0.653 | 1.891(TO7) | 1 − 2 |
| | | | | | 1.932(TO6) | 4 − 5 |
| | C1h | 0.383 | 0.617 | 0.720 | 1.964(TO7) | 1 − 2 |
| | | | | | 1.971(TO7) | 4 − 5 |
| | C2a | 0.369 | 0.529 | 0.619 | 1.892(TO1) | 1 − 2 |
| | | | | | 1.870(TO2) | 4 − 5 |
| | C2b | 0.550 | 1.250 | 1.009 | 1.943(TO2) | 1 − 2 |
| | | | | | 2.011(TO3) | 2 − 3 |
| | | | | | 1.939(TO7) | 4 − 5 |
| | | | | | 2.096(BO23) | |
| | C2c | 0.270 | 0.500 | 0.596 | 1.868(TO2) | 1 − 2 |
| | | | | | 1.870(TO4) | 2 − 3 |
| | | | | | | 3 − 4 |
| | | | | | | 4 - 5 |
| | C2d | 0.435 | 0.564 | 0.679 | 1.911(TO4) | 1 − 2 |
| | | | | | 1.911(TO5) | 4 − 5 |
| | | | | | | 5 − 6 |
| | C2e | 0.437 | 0.531 | 0.685 | 1.884(TO6) | 1 − 2 |
| | | | | | 1.896(TO5) | 4 − 5 |
| | | | | | | 5 − 6 |
| | C2f | 0.360 | 0.512 | 0.666 | 1.911(TO6) | 1 − 2 |
| | | | | | 1.912(TO7) | 4 − 5 |
| | Average | 0.388±0.066 | 0.60±0.19 | 0.693±0.099 | | |

[a] Bonded oxygen atoms in parentheses

Table 3. Copper(II)-ligand interaction energies ($\Delta_{int}E$), Gibbs free interaction energies ($\Delta_{int}G_{298}$) and interaction enthalpies ($\Delta_{int}H_{298}$) at 298 K , their averaged values and standard deviations for the model systems under study.

| Compound | Model | $\Delta_{int}E$ [kJ/mol] | $\Delta_{int}G_{298}$ [kJ/mol] | $\Delta_{int}H_{298}$ [kJ/mol] |
|---|---|---|---|---|
| $^2[Ti_7O_{28}H_{26}Cu]^0$ | A1a | -2586.4 | -2589.3 | -2568.5 |
| | A2a | -2675.8 | -2671.6 | -2635.0 |
| | A2b | -2516.5 | -2517.1 | -2486.4 |
| | Average | -2593±80 | -2593±77 | -2563±74 |
| $^2[Ti_2O_{10}H_{10}Cu]^0$ | B1a | -2695.6 | -2694.7 | -2665.0 |
| | B1b | -2757.2 | -2758.4 | -2726.3 |
| | B1c | -2906.6 | -2903.2 | -2865.4 |
| | B2a | -2696.0 | -2696.7 | -2661.8 |
| | Average | -2764±99 | -2763±98 | -2730±95 |
| $^2[Ti_7O_{30}H_{30}Cu]^0$ | C1a | -2401.9 | -2396.7 | -2343.5 |
| | C1b | -2483.4 | -2476.7 | -2429.0 |
| | C1c | -2387.1 | -2384.3 | -2350.7 |
| | C1d | -2396.9 | -2394.8 | -2346.2 |
| | C1e | -2470.5 | -2467.5 | -2422.4 |
| | C1f | -2453.9 | -2444.9 | -2405.1 |
| | C1g | -2494.4 | -2492.4 | -2445.7 |
| | C1h | -2451.8 | -2454.1 | -2424.6 |
| | C2a | -2453.7 | -2476.0 | -2412.7 |
| | C2b | -2419.1 | -2417.8 | -2381.0 |
| | C2c | -2722.8 | -2720.2 | -2656.1 |
| | C2d | -2453.6 | -2444.4 | -2406.1 |
| | C2e | -2486.5 | -2487.0 | -2457.6 |
| | C2f | -2415.6 | -2410.2 | -2383.8 |
| | Average | -2464±82 | -2462±83 | -2419±77 |

Figure captions:

Chart 1. Ti centers notation in $^{1}[Ti_7O_{28}H_{26}]^{2-}$ - model C

Figure 1.  Optimized structure of $^{1}[Ti_7O_{28}H_{26}]^{2-}$ - model A (Ti - green, O - red, H – gray)

Figure 2.  Optimized structures of $^{1}[Ti_2O_{10}H_{10}]^{2-}$ (top) - model B and $^{1}[Ti_7O_{30}H_{10}]^{2-}$ (bottom) - model C (Ti - green, O - red, H – gray)

Figure 3.  Optimized structures and model notation of $^{2}[Ti_7O_{28}H_{26}Cu]^{0}$ – A1 and A2 model series (Ti - green, O - red, H – gray, Cu - blue)

Figure 4. Optimized structures and model notation of $^{2}[Ti_2O_{10}H_{10}Cu]^{0}$ – B1 and B2 model series (Ti - green, O - red, H – gray, Cu - blue)

Figure 5. Optimized structures and model notation of $^{2}[Ti_7O_{30}H_{30}Cu]^{0}$ – C1 model series (Ti - green, O - red, H – gray, Cu - blue)

Figure 6. Optimized structures and model notation of $^{2}[Ti_7O_{30}H_{30}Cu]^{0}$ – C2 model series (Ti - green, O - red, H – gray, Cu - blue)

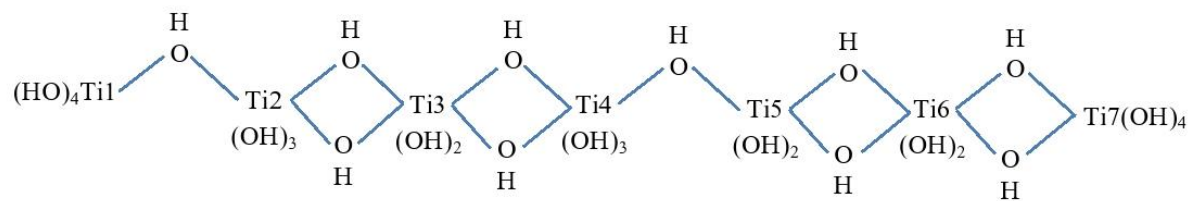

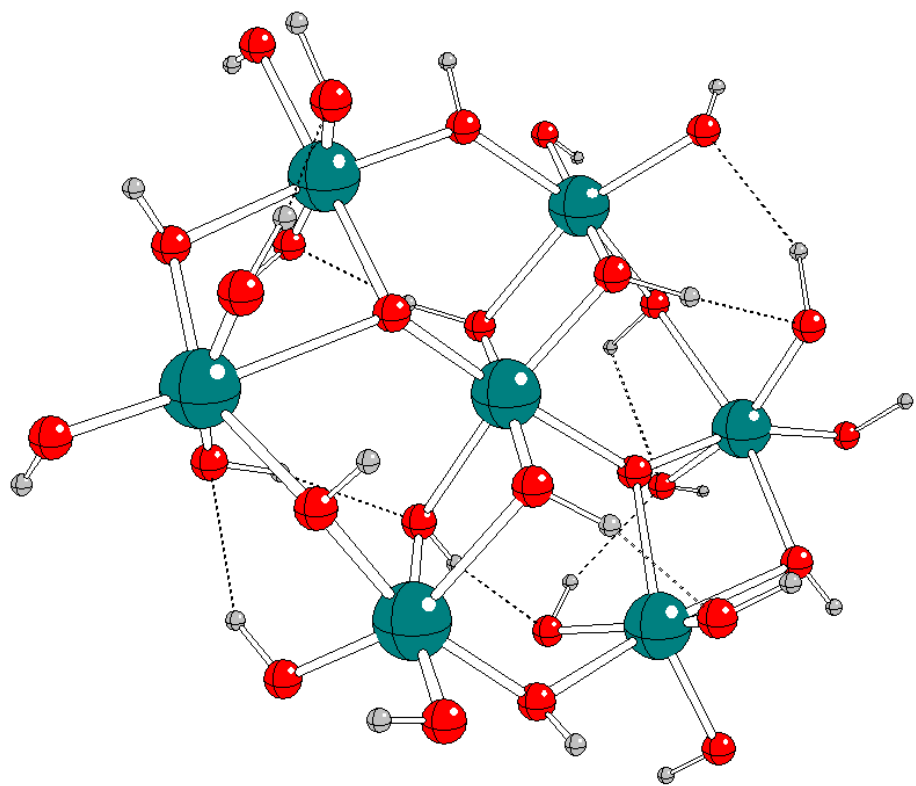

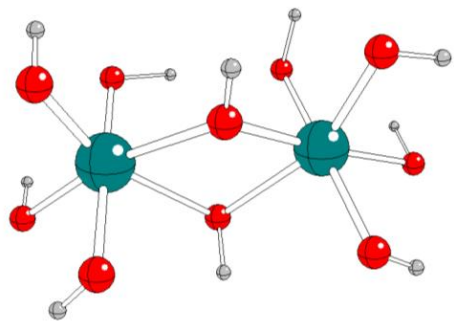

**Model B**

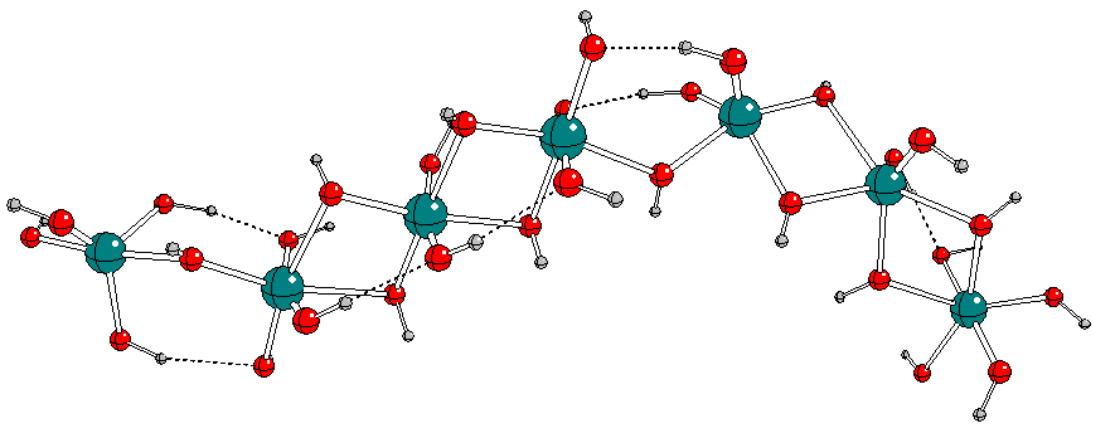

**Model C**

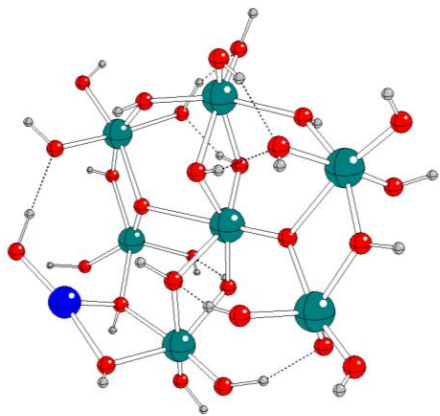

**A1a**

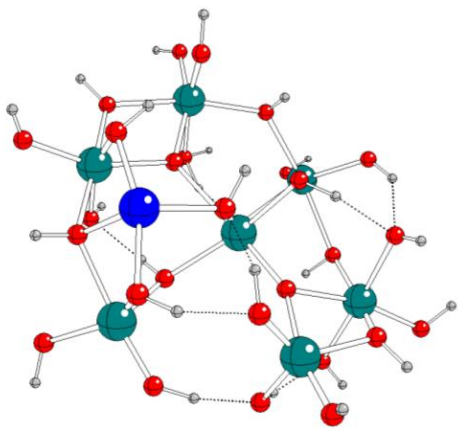

**A2a**

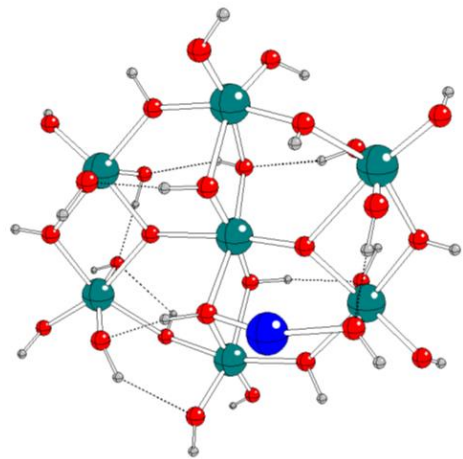

**A2b**

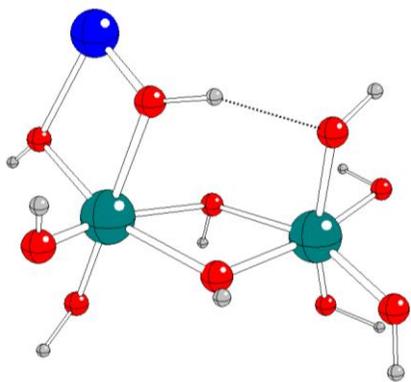
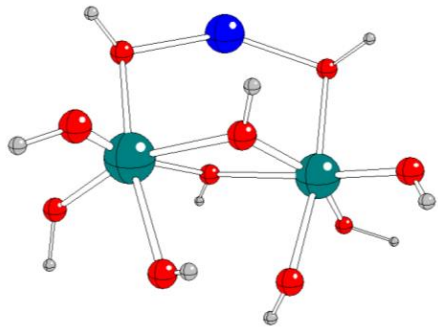

**B1a**          **B1b**

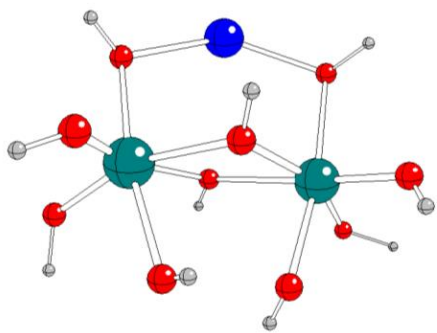
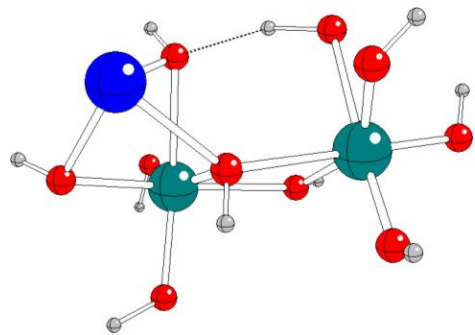

**B1c**          **B2a**

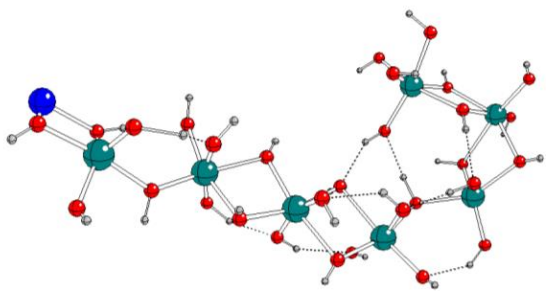

**C1a**

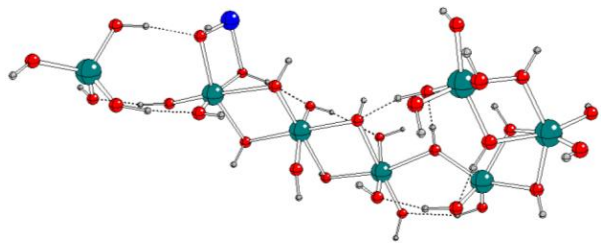

**C1b**

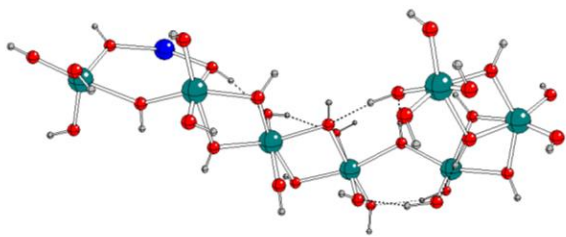

**C1c**

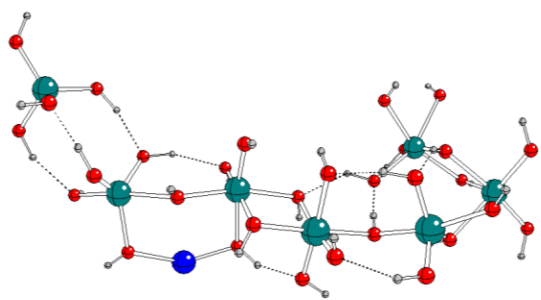

**C1d**

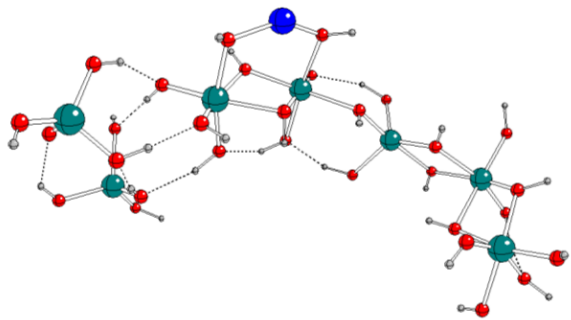

**C1e**

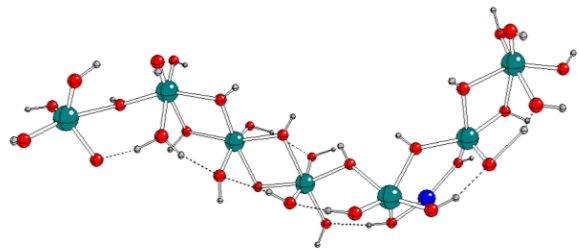

**C1f**

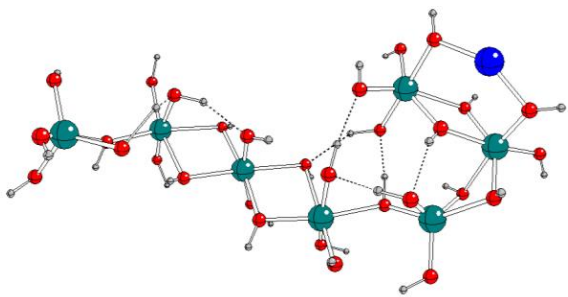

**C1g**

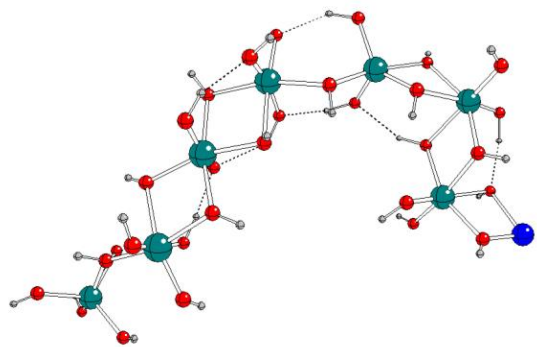

**C1h**

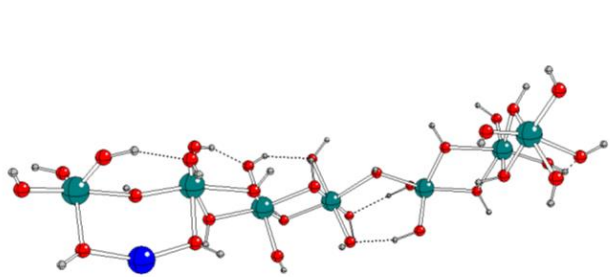

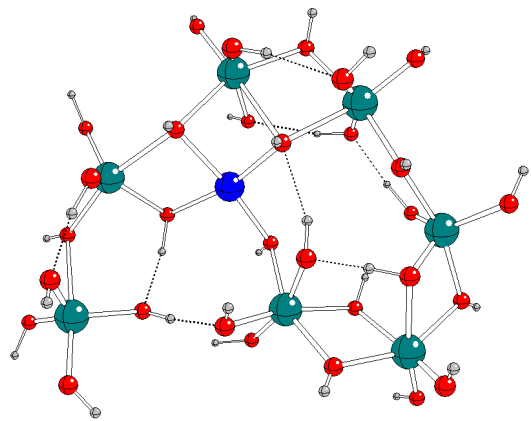

**C2a**

**C2b**

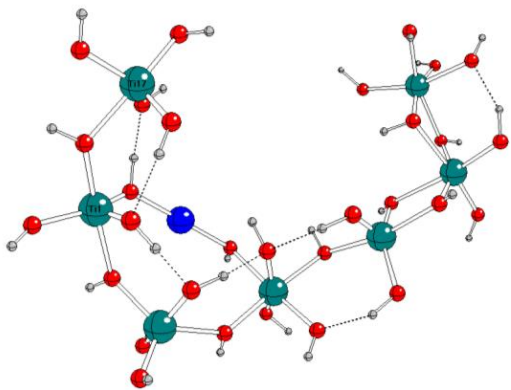

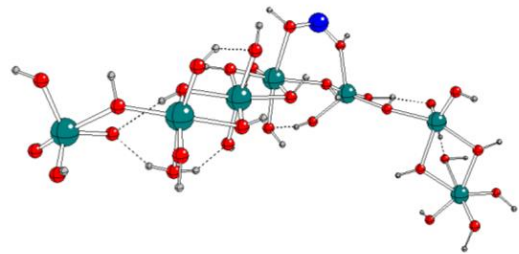

**C2c**

**C2d**

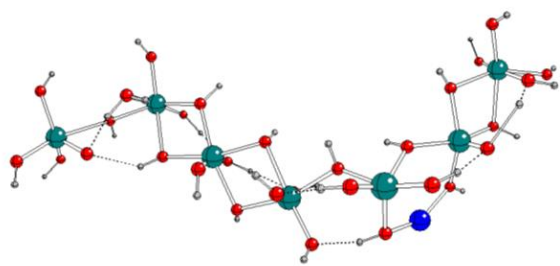

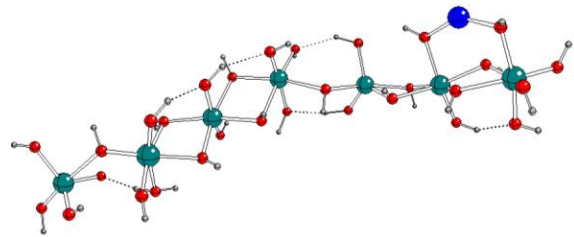

**C2e**

**C2f**